\documentstyle[12pt]{article}
\renewcommand{\theequation}{\mbox{\arabic{section}.\arabic{equation}}}

\def\ii{\'{\char'20}}

\begin{document}

\def\r{\rightarrow}
\def\G{\Gamma}
\def\f{\phi}
\def\p{\partial}
\def\ii{\'{\char'20}}
\def\brr{\begin{array}}
\def\err{\end{array}}
\def\beq{\begin{eqnarray}}
\def\eeq{\end{eqnarray}}
\def\bs{\bigskip}
\def\tr{\mbox{Tr}\, }
\def\ni{\noindent}
\def\wt{\widetilde}
\def\wh{\widehat}
\def\ul{\underline}
\def\nn{\nonumber}
\def\ms{\medskip}
\def\sp{\mbox{Sp}}
\def\re{\mbox{Re}\, }
\def\txs{\textstyle}
\def\dsp{\displaystyle}
\def\ve{V_E({\phi})}
\def\vm{V_{M^4}(\phi)}
\def\wew{\left[\frac{l_1^2}{w_1}+\frac{l_2^2}{w_2}\right]}
\def\bacm{(\Phi_{\mu}^2)}
\def\bam{\Phi_{\mu}}
\def\zm{(z_{\mu}^2)}
\def\la{\lambda}
\def\ins{\int_{\Sigma}d\sigma_x}
\def\cah{{\cal H}}
\def\caz{{\cal Z}}
\def\inb{\int_0^{\beta}d\tau\,\,}
\def\cl{\bar{\Phi}}
\def\act{\bar{S}[\Phi ]}
\def\actcl{\bar{S}[\bar{\Phi}]}
\def\clnorm{|\bar{\Phi}|^2}
\def\clabl{\dot{\Phi}}
\def\fienorm{|\Phi|^2}
\def\actqua{\bar{S}_{ij}[\bar{\Phi}]}
\def\sun{\sum_{n=-\infty}^{\infty}}
\def\sunp{{\sum_{n=-\infty}^{\infty}}\!\!\!^{\prime}}
\def\tint{\int_0^{\infty}dt\,t^{s-1}}
\def\intt{\int_0^{\infty}dt\,}
\newcommand{\bac}{\phi}
\newcommand{\pot}{V(\Phi^2)}
\newcommand{\suz}{\sum_{l_1,l_2=-\infty}^{\infty}}
\newcommand{\reals}{\mbox{${\rm I\!\!R }$}}
\newcommand{\nats}{\mbox{${\rm I\!\!N }$}}
\newcommand{\intgs}{\mbox{${\rm Z\!\!Z }$}}
\newcommand{\komplex}{\mbox{${\rm I\!\!\!\!C }$}}
\newcommand{\ant}{\int\limits}
\newcommand{\sneu}{\sum_{n=1}^{\infty}}
\newcommand{\sleu}{\sum_{l=1}^{\infty}}
\newcommand{\skeu}{\sum_{k=1}^{\infty}}
\newcommand{\snnu}{\sum_{n=0}^{\infty}}
\newcommand{\slnu}{\sum_{l=0}^{\infty}}
\newcommand{\sknu}{\sum_{k=0}^{\infty}}
\newcommand{\snuu}{\sum_{n=-\infty}^{\infty}}
\newcommand{\sluu}{\sum_{l=-\infty}^{\infty}}
\newcommand{\skuu}{\sum_{k=-\infty}^{\infty}}
\newcommand{\cam}{{\cal M}}
\newcommand{\can}{{\cal N}}
\newcommand{\cao}{{\cal O}}
\newcommand{\caf}{{\cal F}}
\newcommand{\mnintx}{\ant_{\cam}d^nx|g(x)|^{\frac 1 2}}
\newcommand{\mvintx}{\ant_{\cam}d^4x|g(x)|^{\frac 1 2}}
\newcommand{\mninty}{\ant_{\cam}d^ny|g(y)|^{\frac 1 2}}
\newcommand{\mdintx}{\ant_{\cam}d^dx|g(y)|^{\frac 1 2}}
\newcommand{\back}{\bar{\Phi}}
\newcommand{\abl}{\partial}
\newcommand{\tintnu}{\ant_0^{\infty}dt}
\newcommand{\th}{\tilde h}
\newcommand{\mnrs}{\mu\nu\rho\sigma}
 \renewcommand{\phi}{\Phi}
\newcommand{\mn}{\mu\nu}
\newcommand{\q}{\left(\xi -\frac 1 6\right)}
\newcommand{\zh}{\zeta_H(s;a)}
\newcommand{\zpi}{2\pi i}
\newcommand{\epa}{E^{\beta}(s)}
\newcommand{\vp}{(4\pi)^2}
\newcommand{\ep}{\epsilon}
\newcommand{\massee}{\left[U''(\Phi )
+\frac{U'(\Phi)}{\Phi}\right]|_{\Phi=\back}}
\newcommand{\massez}{\left[U''(\Phi )-
\frac{U'(\Phi)}{\Phi}\right]^2|_{\Phi=\back}}
\newcommand{\tdh}{\left(\frac 1 {4\pi t}\right)^{\frac 3 2}}

\begin{titlepage}

\title{\begin{flushright}
{\normalsize UB-ECM-PF 94/37 }
\end{flushright}
\vspace{2cm}
{\Large \bf Bose-Einstein condensation for interacting scalar fields
       in curved spacetime
}}

\author{
Klaus Kirsten    \cite{kk}\\
Departament d'ECM, Facultat de F{\ii}sica
\\ Universitat de Barcelona, Av. Diagonal 647, 08028 Barcelona \\
Catalonia, Spain\\
\\and\\
\\
David J. Toms   \cite{djt}          \\
Department of Physics, University of Newcastle Upon Tyne,\\
Newcastle Upon Tyne, United Kingdom NE1 7RU}

\date{January 1995}

\maketitle

\begin{abstract}
We consider the model of self-interacting complex scalar fields with a
rigid gauge invariance under an arbitrary gauge group $G$. In order to
analyze the phenomenon of Bose-Einstein condensation finite temperature
and the possibility of a finite background charge is included. Different
approaches to derive the relevant high-temperature behaviour of the
theory are presented.
\end{abstract}
\end{titlepage}

\section{Introduction}
\setcounter{equation}{0}

Although Bose-Einstein condensation is well-known to occur for non-relativistic
spin-0 bosons at low temperatures, it is only comparatively recently that the
analogous phenomenon was studied in relativistic quantum field theory
\cite{haberprl,haber,kapusta,bernstein,benson}. For relativistic fields
 Bose-Einstein condensation occurs at high
temperatures and can be interpreted in terms of spontaneous symmetry breaking.

The extension and generalization of Bose-Einstein condensation to curved
spacetimes, and to spacetimes with boundaries has also been the subject of much
study. The non-relativistic Bose gas in the Einstein static universe
was treated
by Altaie \cite{Altaie}. The generalization to relativistic scalar fields was
given in Refs. \cite{singh,parkerzhang}. An extension to higher
dimensional
 spheres was given
by Shiraishi \cite{Shiraishi}. More recently a detailed critical
examination of
Bose-Einstein condensation in the static Einstein universe has been
performed
\cite{smithtoms}. Bose-Einstein condensation on hyperbolic manifolds
 \cite{cognola},
and in the Taub universe \cite{huang} has also been considered.

By calculating the high temperature expansion of the thermodynamic
potential
when boundaries are present, Ref.~\cite{doktor} was able to examine
Bose-Einstein
condensation in certain cases. Later work \cite{bose} showed how to
interpret
Bose-Einstein condensation in terms of symmetry breaking in the manner
of the
flat spacetime calculations \cite{haber,kapusta}. A crucial feature
when the
spacetime is
curved, or when boundaries are present, is that if the symmetry breaks
then the
ground state cannot be constant in general \cite{bose}. A later study
\cite{diag} showed how interacting scalar fields could be treated.

We want to continue the considerations on the Bose-Einstein condensation
taking as a model a set of interacting scalar fields with a rigid gauge
invariance under the action of an arbitrary gauge group $G$. In order to
examine the phenomenon of Bose-Einstein condensation in relativistic
quantum field theories, the interesting regime is the high-temperature
one to which we will concentrate from the beginning. In doing that,
different approaches may be developed and we will present several of
them at different stages of the paper.

For explicity and in order to develop the zeta function approach in
combination with heat-kernel techniques, we concentrate first on the
$O(N)$ self-interacting Bose gas in curved spacetime at finite
temperature and finite chemical potentials.
This continues the analysis started in \cite{haberprl,haber,kapusta}
from flat to curved spacetime. In section 2 we describe in some detail
the model and derive briefly the equations necessary for the following
considerations. The aim of section 3 is to derive the high-temperature
behaviour of the theory using an expansion around the free field theory.
The leading
dependence on the chemical potentials $\mu_{\alpha}$ will be given by
the one of the free field. The leading term due to interaction may be
determined by considering the theory with $\mu_{\alpha} =0$ (this may be
seen by dimensional analysis). For that reason,
in section 3 we derive the high-temperature behaviour of the
Bose gas for vanishing chemical potential and for arbitrary $N$. In that
case it is no additional complication to allow for nonconstant
background field and we will consider this general case.
Afterwards, in section 4, we concentrate on the $O(2)$ model and we
restrict our attention to a constant background field. Because we are
only interested in the leading dependence of the effective action on the
temperature, this is actually no restriction because by dimensional
grounds it is clear that derivatives of the background field will not
enter. This has however the advantage that calculations are much more
explicit in that for example an explicit knowledge of the excitation
energies is given. We will develop two different approaches. The first
one is once more based on the zeta function approach, the second one is
based on a weak field expansion of the excitation energies. In both
cases we determine the leading two terms in the high temperature
expansion. Section 5 contains remarks on the limit
$N\to \infty$ in the $O(N)$ model, especially the influence of
higher loop corrections is briefly discussed. In section 6 we extend our
results to general
gauge groups. Here we choose the approach based on the excitation
energies. In that part the chemical potential is treated as a small
interaction term and we use normal quantum-field perturbation theory. In
this approach and for the general gauge group we indicate how results
containing higher orders in the temperature expansion may be obtained
and how a nonconstant background field may be treated with in a
systematic way.

\section{Incorporation of $\mu$ in an $O(N)$ model of a self-interacting
Bose gas}
\setcounter{equation}{0}
In this section we will derive the basic equations of a model of
$N$ real scalar fields with an arbitrary self-interaction coupled to a
classical gravitational field. We are
especially interested in the influence that conserved background charges
have on the theory. As has been argued in detail \cite{haber}, the
maximum number of mutually commuting charges is $N/2$ for $N$ even and
for definiteness we will always take $N$ even.

We will consider the $O(N)$ model in an
arbitrary ultrastatic spacetime \cite{fulling} with
\begin{equation}
          ds^2=dt^2-g_{ij}({\bf x})dx^idx^j\;.\label{eq:1.1}
\end{equation}
$x^i$ are local coordinates on the spatial hypersurface $\Sigma$ which
is taken
to be a $D$-dimensional Riemannian with a finite volume $V$. The
infinite volume limit can be taken
at the end. The scalar field action is
\begin{equation}
          S=\int dt\int_{\Sigma}d\sigma_x
                         \Big\lbrace
\frac 1 2 (\partial^\mu\Phi_a)
               (\partial_\mu\Phi_a) -\frac 1 2 (m^2-\xi
 R)|\Phi|^2-V_{tree}(\Phi)\Big
               \rbrace,\label{eq:1.2}
\end{equation}
where $V_{tree}(\Phi)$ is the interaction potential at the tree
level. We
will
assume that this potential only depends on $\Phi_a$
through
$|\Phi|^2=\Phi_a\Phi_a$. Here and in the
following the summation over $a=1,...,N$ is always included.
$d\sigma_x=\sqrt{g_{ij}(\bf x)}d^Dx$ is the volume element on $\Sigma$.

Associated with the invariance of (\ref{eq:1.2}) under global gauge
transformations, there are $N(N-1)/2$ conserved currents
$J_{ab}^{\mu}$ giving rise to the conserved charges
$Q_{ab}$~: \begin{equation}
          Q_{ab}=\int_{\Sigma}d\sigma_x\,J_{ab}^0\label{eq:1.3}
\end{equation}
where
\begin{equation}
J_{ab}^0=i\Big(\Phi_a\dot{\Phi}_b-\Phi_b
\dot{\Phi}_a\Big)\;.
          \label{eq:1.4}
\end{equation}
A quantum state of the system is specified by the eigenvalues of a
complete set of mutually commuting charges. As mentioned (see
\cite{haber}), the maximum number of mutually commuting charges is
$N/2$. A convenient set of mutually commuting charges is
$Q_{12},Q_{34},...$, which we label by
\beq
Q_{\alpha} =Q_{2\alpha-1,2\alpha} =  \ins j_{\alpha}^0.
\label{eq:1.5}
\eeq

The conservation of charges is implemented in the usual way by
introducing
Lagrange multipliers $\mu_{\alpha}$ and defining the generalized
Hamiltonian density $\bar{{\cal H}}$ by
\begin{equation}
\bar{{\cal H}}={\cal H}-\mu_{\alpha} J_{\alpha}^0,\label{eq:1.6}
\end{equation}
where ${\cal H}$ is the usual Hamiltonian density.
Here and in the following, greek indices are always summed from $1$ to
$N/2$.

The canonical momenta which follow from (\ref{eq:1.2}) are
\begin{equation}
          \Pi_a=\dot{\Phi}_a,
                   \label{eq:1.7}
\end{equation}
and we have
\begin{equation}
          {\cal
H}=\frac 1 2 \Pi_a\Pi_a+\frac 1 2 |\nabla\Phi|^2+
                     \frac 1 2  (m^2+\xi R)|\Phi|^2+
               V_{tree}(\Phi )\;.\label{eq:1.8}
\end{equation}
The generalized Hamiltonian density follows simply from
 (\ref{eq:1.4})-(\ref{eq:1.8}).
Here
$|\nabla\Phi|^2=g^{ij}(\partial_i\Phi_a)
(\partial_j\Phi_a)$.

The reason for adopting a Hamiltonian approach is, that this is the
easiest way to incorporate finite temperature effects in quantum field
theory. First of all perform a Wick rotation $\tau =it$ to
obtain a $(D+1)$-dimensional Riemannian spacetime. Since we have
restricted the spacetime to be ultrastatic, this presents no difficulty.
The grand partition function is expressed as
\beq
\caz =\int [d\Pi_a][d\Phi_a]\left[\inb\ins
    \left\{i\Pi_a \clabl _a -\bar{\cah}\right\}\right].
\label{eq:1.9}
\eeq
The path integral in (\ref{eq:1.9}) extends over all fields $\Phi_a$
periodic in time with period $\beta =1/T$. Because $\bar{\cah}$ is
quadratic in the momenta, the integration over the momenta in
(\ref{eq:1.9}) may be performed simply by completing the square
\cite{bernard}. This leaves a configuration space path integral
\beq
\caz =\int [d\Phi_a]\exp (-\act ),\label{eq:1.10}
\eeq
where
\beq
\act &=& \inb\ins \left\{\frac 1 2
   \left(\clabl _{2\alpha -1} -i \mu_{\alpha}
     \Phi_{2\alpha}\right)^2
+  \frac 1 2 \left(\clabl _{2\alpha} +i \mu_{\alpha}
   \Phi_{2\alpha-1}\right)^2\right.\nn\\
& & \left.-\frac 1 2 |\nabla \Phi|^2 -\frac 1 2 (m^2 +\xi R)\fienorm
    -V_{tree} (\Phi )\right\} . \label{eq:1.11}
\eeq
Rather than deal with the thermodynamic potential, we will consider the
finite temperature effective action computed using the background field
method \cite{background}. Thus we introduce some scalar background field
$\cl$. By expanding about this background in the usual way, the one-loop
effective action turns out to be
\beq
\Gamma &=&\actcl +\frac 1 2 \ln\det\left\{\lambda^2 \actqua
\right\}\nn\\ & =&\actcl +\Gamma ^{(1)},
\label{eq:1.12}
\eeq
where
\beq
\actqua =\frac{\delta^2 \actcl }{\delta \cl _i(x) \delta \cl _j (x')}.
\label{eq:1.13}
\eeq
The second term $\Gamma ^{(1)}$ in (\ref{eq:1.12}) arises from the
Gaussian functional
integration, and represents the one-loop quantum correction to the
classical action functional. $\lambda$ is a unit of length introduced to
keep the argument of the logarithm dimensionless. Using (\ref{eq:1.11}),
for $\actqua$ we find
\beq
\actqua =(M_{i,j}+V_{ij})\delta (x,x'),\label{eq:1.14}
\eeq
where
\beq
\left(\begin{array}{cc}
     M_{2\alpha -1,2\alpha -1}    &     M_{2\alpha -1,2\alpha}\\
     M_{2\alpha,2\alpha -1}    &     M_{2\alpha ,2\alpha}
      \end{array}\right)   =
\left(\begin{array}{cc}
     -\Box -\mu_{\alpha}^2 +m^2 +\xi R   &     2i \mu_{\alpha}
                               \frac{\partial}{\partial\tau}\\
     - 2i\mu_{\alpha}\frac{\partial}{\partial\tau}  &
                           -\Box -\mu_{\alpha}^2 +m^2 +\xi R
      \end{array}\right),  \label{eq:1.15}
\eeq
the other components are zero, and
\beq
V_{ij} =V''_{tree} (\cl )e_ie_j +\frac{V'_{tree}
(\cl )}{\cl}
         (\delta_{ij} -e_ie_j).\label{eq:1.16}
\eeq
Here we introduced $e_i =\cl _i/\clnorm$. The separation in $M$ and $V$
is a separation in free field plus terms arising due to interaction.

In order to calculate the functional determinant of the operator
$\actqua$, we will use the zeta function definition introduced by
Hawking \cite{haw} and by Critchley, Dowker \cite{critch}. Then
(\ref{eq:1.12}) reads
\beq
\Gamma =\actcl -\frac 1 2 \left[\zeta ' (0) +\zeta (0) \ln\lambda
^2\right],\label{eq:1.17}
\eeq
where $\zeta (s)$ is the zeta function associated with the operator
$\actqua$, equation (\ref{eq:1.14}).

We are especially interested in the high temperature behaviour of the
theory. In that regime heat-kernel techniques have been shown to be a
very powerful tool \cite{heat,bose} and will also be used here. In doing
that, one of the encountered problems is the diagonalization of
$\actqua$ \cite{diag,onmodel}. For $\mu_{\alpha} =0$ this may be done
\cite{onmodel} and we will first consider this case in detail.

\section{$O(N)$-model with vanishing chemical potentials}
\setcounter{equation}{0}
In considering vanishing chemical potentials $\mu_{\alpha} =0$, the
problem is enormously simplified, because the operator $M$,
(\ref{eq:1.15}), is diagonal, $M_{ij} =-\delta_{ij}(\Box +\xi R +m^2 )$.
For that case,
the eigenvalues of the operator $\actqua$ may be represented in the form
\beq
\nu_{n,j} =\left(\frac{2\pi}{\beta}\right)^2 n^2+\lambda_j,
\label{eq:2.1}
\eeq
with the eigenvalues $\lambda_j$ of the operator
\beq
D= -\Delta +U(\cl ).\label{eq:2.2}
\eeq
For simplicity we introduced here
\beq
U_{ij} (\cl ) =V_{ij} +(\xi R +m^2 )\delta_{ij}.\nn
\eeq
Using a Mellin transformation, the zeta function of $\actqua$ may be
written as
\beq
\zeta (s) &=& \frac 1 {\Gamma (s)}
   \sun \sum_j \tint \exp\left\{-\left[\left(\frac{2\pi}{\beta}
    \right)^2 n^2 +\lambda_j \right] t \right\}\nn\\
  &=&
 \frac 1 {\Gamma (s)}
   \sun \tint \exp\left\{-\left(\frac{2\pi}{\beta}
                  \right)^2 n^2 t\right\} \tr K(x,x,t),
        \label{eq:2.3}
\eeq
where the kernel $K(x,x',t)$ satisfies the equation
\beq
-\frac{\partial}{\partial t}K(x,x',t) &=& (-\Delta +U(\cl
))K(x,x',t),\nn\\
\lim_{t\to 0} K(x,x',t) &=& |g|^{-\frac 1 2}\delta (x,x').
\label{eq:2.4}
\eeq
We would like to concentrate on the high-temperature behaviour of the
theory. Then for $n\neq 0$ only small values of the parameter $t$
contribute considerably to the integral in (\ref{eq:2.3}). So it is
reasonable for $n\neq 0$ to make use of the asymptotic behaviour for
$t\to 0$ of $K(x,x',t)$ \cite{kernel}. As mentioned, in this section we
will not assume that the background field $\cl $ is a constant. However,
in order that the expansion that we are going to derive is reasonable,
$\cl $ has to be slowly varying, together with the condition $\beta
|R|^{\frac 1 2} \ll 1$, where $|R|$ is the magnitude of a typical
curvature of the spacetime.

The suitable tool to derive the described expansion is the use of the
following ansatz for the heat-kernel suggested by Parker and Toms
\cite{david},
\beq
K(x,x',t) &=& \left(\frac 1 {4\pi t}\right)^{\frac D 2}
             \exp\left\{-\frac{\sigma (x,x')}{2t} -\left[
              U(\cl )-\frac 1 6 R\right] t\right\}\times\nn\\
   & & \hspace{1cm} \Delta_{VM}(x,x') \Omega (x,x',t),
\label{eq:2.5}
\eeq
where $2\sigma (x,x')$ is the square of the proper arc length along the
geodesic from $x'$ to $x$ and $\Delta _{VM}(x,x')$ is the Van
Vleck-Morette determinant. For $t\to 0$ the function $\Omega (x,x',t)$
may be expanded in an asymptotic series,
\beq
\Omega (x,x',t) =\sum_{l=0}^{\infty} a_l (x,x') t^l , \label{eq:2.6}
\eeq
where the coefficients have to fulfill some recurrence relation. Using
the given ansatz, (\ref{eq:2.5}), it has been shown by Jack and Parker
\cite{leonard}, that the dependence of $a_l$, $l=1,...,\infty$, on the
field $\cl $ is only through derivatives of the field. As a result,
combining (\ref{eq:2.3}), (\ref{eq:2.5}) and (\ref{eq:2.6}), we will
automatically find a derivative expansion in the field $\cl $ without
any effort.

Although the method in principle works to every order, it gets very
cumbersome after the leading terms. For that reason we concentrate from
the beginning only on the leading terms. However, the method will be
clear after that.

First of all one has
\beq
\zeta (s) =\zeta_{\Sigma} (s) +\zeta_z (s),\label{eq:2.7}
\eeq
with the zeta function $\zeta_{\Sigma}$ of the spatial section,
\beq
\zeta_{\Sigma} (x) =\frac 1 {\Gamma (s)}\tint \tr K(x,x,t),
\label{eq:2.8}
\eeq
resulting from the $n=0$ term, and the finite temperature part
\beq
\zeta_z (s) =\frac 1 {\Gamma (s)} \sunp \,\,\tint \exp\left\{
        -\left(\frac{2\pi }{\beta}\right)^2n^2t\right\}
\tr K(x,x,t),\label{eq:2.9}
\eeq
where the prime indicates that the $n=0$ term of the sum is omitted. The
zero temperature contribution represented by (\ref{eq:2.8}) has been
analyzed in detail in \cite{onmodel} using an adiabatic expansion
method, so let us concentrate on
the finite temperature part, eq.~(\ref{eq:2.9}). Use of the heat-kernel
expansion (\ref{eq:2.5}) leads to
\beq
\zeta _z(s) =\sum_{i=0}^{\infty} \zeta_z^i (s)\label{eq:2.10}
\eeq
with
\beq
\zeta_z^i (s) &=&
    \frac 1 {(4\pi)^{\frac D 2} \Gamma(s)}
     \sunp \,\,\intt t^{s-1-\frac D
2}\exp\left\{-\left(\frac{2\pi}{\beta}
    \right)^2n^2 t\right\}\nn\\
  & & \hspace{2cm}\tr\left\{\exp\left[-( U(\cl )-\frac 1 6 R
)t\right]a_it^i
          \right\}.\label{eq:2.11}
\eeq
We will concentrate only on the first two terms, $\zeta _z^0 (s)$ and
$\zeta _z ^2 (s)$ ($\zeta _z^1 (s)$ is zero due to $a_1 =0$), containing
the leading terms of the high-temperature expansion. As one realizes, a
diagonalization of the potential matrix $U (\cl )$ is needed. For
completeness, the matrix $S$ accomplishing the diagonalization is
presented in the Appendix A (for details see \cite{onmodel}). Using the
results presented there, one first finds
\beq
\zeta_z ^0 (s) &=& \frac 2 {(4\pi )^{\frac D 2} \Gamma (s)}
\sum_{n=1}^{\infty} \intt t^{s-1-\frac D 2}\exp\left\{ -
      \left(\frac{2\pi }{\beta}\right)^2 n^2 t\right\}\times\nn\\
    & &\hspace{1cm}\left[\exp\{-M_1^2 t\} +(N-1) \exp\{-M_2^2
t\}\right], \label{eq:2.12}
\eeq
where we introduced the effective masses
\beq
M_1^2 &=& m^2 +\left(\xi -\frac 1 6 \right) R +V_{tree}'' (\cl ),
\label{eq:2.13}\\
M_2^2 &=& m^2 +\left(\xi -\frac 1 6 \right) R + \frac{V_{tree}' (\cl
)}{\cl}. \label{eq:2.14}
\eeq
These masses, or more detailed $(M_i\beta)^2$, play the role of a
natural expansion parameter of the theory. Performing the integration in
(\ref{eq:2.12}) leads to Epstein type zeta functions,
\beq
E_1^{c^2} &=& \sum_{l=1}^{\infty} [l^2+c^2]^{-\nu}\nn\\
   &=& \sum_{l=0}^{\infty} (-1)^l\frac {\Gamma (s+l)}{l!\Gamma (s)}
        \zeta_R (2s+2l) c^{2l}, \label{eq:2.15}
\eeq
where the last equality holds for $|c^2| <1$ \cite{klaus92}. $\zeta _R
(s)$ is the Riemann zeta function. With this, in principle all terms in
the high temperature expansion may be calculated.

However, for simplicity let us present only the case $D=3$ and only the
three leading terms. The relevant results then read
\beq
\zeta_z ^0 (0) =\frac {\beta}{2(4\pi )^2} \ins [M_1^4+(N-1)M_2^4]
       \label{eq:2.16}
\eeq
and
\beq
{\zeta_z'}^0 (0) &=& \frac 1 {(4\pi) ^{\frac 3 2}}\ins
     \left\{\frac{\sqrt{\pi}}{45}N\left(\frac{2\pi}{\beta}\right)^3
        -\frac{\sqrt{\pi}} 3 \left(\frac{2\pi}{\beta}\right) [M_1^2
                    +(N-1)M_2^2]\right.\nn\\
   & &\hspace{1cm}  +\frac{\beta}{2\sqrt{\pi}} \left[\gamma
+\ln\left(\frac{\beta}{4\pi}\right)\right]
    [M_1^4+(N-1)M_2^4]\nn\\
& &\left.\hspace{1cm} {\cal O} ((M_i\beta)^6)\right\}.\label{eq:2.17}
\eeq
For $\zeta_z ^2 (s)$ we continue as indicated in the Appendix A,
eq.~(\ref{eq:a.3}). Including terms only up to $\beta^{-3} {\cal O}
((M_i\beta )^4)$, the result reads
\beq
\zeta _z^2 (s) =\frac 2 {(4\pi)^{\frac 3 2}} \frac{\Gamma
\left(s+\frac
1 2\right)}{\Gamma (s)}\left(\frac{\beta}{2\pi}\right)^{2s+1}
\zeta_R (2s+1) \ins A\label{eq:2.18}
\eeq
with
\beq
A&=& \frac N {180} (R_{\mu\nu\rho\sigma} R^{\mu\nu\rho\sigma}
        -R_{\mu\nu}R^{\mu\nu} )-\frac N 6 \left(\xi -\frac 1 5 \right)
       \Box R\nn\\
& & -\frac{N-1} 6 \Box \left(\frac{V'_{tree}(\cl )}{\cl }\right)
     -\frac 1 6 \Box V_{tree}'' (\cl ) .\label{eq:2.19}
\eeq
This gives the contributions
\beq
\zeta _z^2 (0) =\frac 1 {2(4\pi )^{\frac 3 2}}\frac{\beta}
             {\sqrt{\pi}}\ins A\label{eq:2.20}
\eeq
and
\beq
{\zeta _z'}^2 (0) = \frac 1 {(4\pi )^{\frac 3 2}}\frac{\beta}
{\sqrt{\pi}} \left[\gamma +\ln\left(\frac{\beta}{4\pi}\right)\right]
      \ins A.\label{eq:2.21}
\eeq
Dealing with a manifold without boundary, the terms in $A$ containing
the Laplace-Beltrami operator vanish.

Putting things together, we find for the high-temperature behaviour of
the $O(N)$-model the result
\beq
\Gamma^{(1)}&=&-\frac 1 {2(4\pi)^{\frac 3 2}}\ins\left\{
    \frac{\sqrt{\pi}}{45} N\left(\frac{2\pi}{\beta}\right)^3
       -\frac{\sqrt{\pi}} 3 \frac{2\pi}{\beta} [M_1^2
+(N-1)M_2^2]\right.\nn\\
& &\hspace{1cm} +\frac{\beta}{2\sqrt{\pi}}\left[\gamma
     +\ln\left(\frac{\beta\lambda}{4\pi}\right)\right]
      [M_1^4+(N-1)M_2^4+2A]\nn\\
& &\hspace{1cm} \left.+\beta^{-3} {\cal O} ((M_i
\beta)^6)\right\}     -\frac 1 2 \zeta'_{\Sigma } (0).
\label{eq:2.22} \eeq
For the case $N=1$ it reduces to the results found in \cite{finite}, as
it should.

Let us now consider the case $\mu _{\alpha} \neq 0$. Because
we will indicate how to develop a systematic approach in section 6 for
the case of a general gauge group, here only some comments. Starting
point of the calculation
is \beq \zeta (s) =\frac 1 {\Gamma (s)}\sun \tint \tr\left\{\exp[\actqua
t]\right\}\label{eq:2.23}
\eeq
with $\actqua$ given in (\ref{eq:1.14}). As seen in (\ref{eq:1.15}) and
(\ref{eq:1.16}), $\actqua$ splits into the free case and, for small
coupling, into small corrections due to the self-interaction.
Treating the self-interaction using normal quantum-field perturbation
theory, the leading term clearly gives the theory of $N/2$ charged
scalar fields with associated chemical potentials $\mu_{\alpha}$
and
the effective potential, or up to the order we write it down also the
thermodynamic potential, for the $O(N)$ model at finite $T$ and finite
$\mu_{\alpha}$ reads,
\beq
\Omega &=& -\frac{\pi^2}{90} NVT^4 -\frac 1 6 T^2 V\sum_{\alpha
=1}^{N/2}
   \mu_{\alpha}^2 \nn\\
& &+\frac 1 {24} T^2\ins [M_1^2 +(N-1) M_2^2]+...\label{eq:2.25}
\eeq
In section 6 we will generalize this formula from the gauge group $O(N)$
to an arbitrary rigid gauge group $G$.
\section{$O(2)$ model at finite chemical potential}
\setcounter{equation}{0}
Let us now concentrate on the $O(2)$ model with a constant classical
background field. For $N=2$ the analysis is somehow more explicit and
systematic approaches may be
developed at finite background charge. We will present two of them.
One is based on an expansion around $\mu =0$, the second one is an
expansion in the coupling constant which is assumed to be small, this is
in powers of the potential $V_{tree}$ and its derivatives, as indicated
at the end of section III.
In both cases
dimensional analysis shows, that no other type of contributions may
appear and that the expansion up to a certain order is complete.
In this section we will also include the possibility that the manifold
has a boundary.
\subsection{Exansion around $\mu =0$}
One possibility to treat the
$O(2)$ model in curved spacetime is to follow the approach
developed for the theory in the Minkowki spacetime \cite{benson}. In
detail the fluctuation operator is given by
\beq
\bar S_{11}&=&\delta^4(x-y)\left[-\Box -\mu ^2+U''(\phi)\frac{\phi_1^2}
{\phi^2}+\frac{U'(\phi)}{\phi}\frac{\phi_2^2}{\phi^2}\right]
_{\phi=\back},\nn\\
\bar S_{12}&=&\delta^4(x-y)\left[2i\mu\frac{\abl}{\abl\tau}
+U''(\phi)\frac{\phi_1\phi_2}
{\phi^2}-\frac{U'(\phi)}{\phi}\frac{\phi_1\phi_2}{\phi^2}\right]
_{\phi=\back},\nn\\
\bar S_{21}&=&\delta^4(x-y)\left[-2i\mu\frac{\abl}{\abl\tau}
+U''(\phi)\frac{\phi_1\phi_2}
{\phi^2}-\frac{U'(\phi)}{\phi}\frac{\phi_1\phi_2}{\phi^2}\right]
_{\phi=\back},\label{25a}\\
\bar S_{22}&=&\delta^4(x-y)\left[-\Box -\mu^2+U''(\phi)\frac{\phi_2^2}
{\phi^2}+\frac{U'(\phi)}{\phi}\frac{\phi_1^2}{\phi^2}\right]
_{\phi=\back},\nn
\eeq
with $U(\cl ) =(1/2)(m^2+\xi R)\cl ^2 +V_{tree}(\cl )$.

Doing the Gaussian integration, the one-loop approximation of the
partition sum is
\beq
\ln\caz^{(1)}&=&-\frac 1 2 \mbox t
\mbox r\ln\left\{\Box^2-4\mu^2\abl_{\tau}^2
-\left(U''(\phi)+\frac{U'(\phi)}{\phi}-2\mu^2\right)\Box\right.\label{26}
\\ & &\left.\hspace{2cm}+\left(U''(\phi)-\mu^2\right)\left(
\frac{U'(\phi)}{\phi}-\mu^2\right)\right\}_{\phi=\back}.\nn
\eeq
Introducing
\beq
\gamma^2=\left(-\Delta +U''(\back)-\mu^2\right)\left(-\Delta
+\frac{U'(\back)}{\back}-\mu^2\right)\label{27}
\eeq
and
\beq
\alpha^2=2\left[-\Delta +\mu^2+\frac 1 2
\left(U''(\back)+\frac{U'(\back)}{\back}
\right)-\gamma\right],\label{28} \eeq
this may be rewritten as
\beq
\ln\caz^{(1)}=\frac 1 2
\mbox t \mbox r \ln\left\{(-\abl_{\tau}^2+\alpha\abl_{\tau}+\gamma)
(-\abl_{\tau}^2-\alpha\abl_{\tau}+\gamma)\right\}.\label{29}
\eeq
Assuming now a complete basis of eigenfunctions of the Laplace-Beltrami
operator $-\Delta$ with eigenvalues $\lambda_n$ and denoting the
resulting quantities of eqs.~(\ref{27}), (\ref{28}), as $\gamma_n^2$ and
$\alpha_n^2$, this results formally in
\beq
\ln\caz^{(1)}=E_{zero-point}+\Omega,\label{210}
\eeq
where we introduced the infinite zero-point energy,
\beq
E_{zero-point}=\sum_j\frac{E_{+,j}+E_{-,j}} 2,\label{211}
\eeq
and the thermodynamic potential
\beq
\Omega=\frac 1 {\beta}\sum_j\left[\ln\left( 1-e^{-\beta
E_{+,j}}\right)+\ln\left( 1-e^{-\beta
E_{-,j}}\right)\right].\label{212}
\eeq
Furthermore, the excitation energies are
\beq
E_{\pm ,n}=\sqrt{\gamma_n+\frac{\alpha_n^2} 4}\pm\frac{\alpha_n}
2 .\label{213}
\eeq
The representation eqs.~(\ref{211}), (\ref{212}), of the one-loop grand
partition
sum is completely analogous to the case of a free scalar field in curved
spacetime, the only change is in the excitation energies. Putting
$U(\phi)=(1/2)(m^2+\xi R)\phi^2$, this is the free field, the grand
thermodynamic potential of a free field propagating in curved background
is thus found directly.

Also for $N=2$,
in general it will be impossible or at least very difficult to treat the
one-loop contribution, eq.~(\ref{210}), exactly. However, we are
interested in the Bose-Einstein condensation and the relevant
range for this phenomenon
is the high-temperature regime to which we will once more
concentrate.

By the mentioned analogy with the free field, one may obtain without any
calculation the high-temperature expansion in terms of quantities given
through the zeta-function of the excitation energies,
\beq
\zeta_{\pm}(\nu)=\sum_j (E_{\pm ,j}^2)^{-\nu},\label{31}
\eeq
or the associated heat-kernel coefficients,
\beq
K_{\pm}(t)=\sum_j e^{-E_{\pm ,j}^2 t}\sim\left(\frac 1 {4\pi
t}\right)^{\frac 3 2} \sum_{l=0,1/2,...}^{\infty}b_l^{\pm}t^l.\label{32}
\eeq
By introducing $E_{\pm ,zeropoint}$, $\Omega_{\pm}$, as
the
quantities in (\ref{211}) and (\ref{212}) resulting from the excitation
energies $E_{\pm, j}$, the results read
\beq
\Omega_{\pm} &=&-\frac 1 {2}PP\zeta_{\pm}(-\frac 1 2 )+\frac 1
{(4\pi)^2}\label{33}\\
& &\times\left\{-b^{\pm}_2
\ln\left(\frac{\beta}{2\pi}\right)+\psi (2)+
\frac{2\sqrt{\pi}}{\beta}b_{\frac 3 2}^{\pm}\ln\beta
+P_{\pm}+S_{\pm}\right\},\nn \eeq
with
\beq
S_{\pm}=-\sum_{r=1/2,1,...}^{\infty}b_{2
+r}^{\pm}\left(\frac{\beta}{4\pi}\right)^{2r}\frac{(2r)!}{\Gamma(r+1)}\zeta_R
(1+2r)\label{34}
\eeq
and
\beq
P_{\pm}=-\sum_{r=0,1/2,1}b^{\pm}_r\left(\frac{\beta}2\right)^
{-4+2r} \Gamma\left(2 -r\right)\zeta_R(4-2r).\label{35}
\eeq
Here, $PP\zeta_{\pm}(-1/2)$ denotes the finite part of
$\zeta_{\pm}(s)$ at $s=-1/2$.

Thus we have reduced the
task of determining the high-temperature behaviour of the theory to an
analytical treatment of $\zeta_{\pm}(\nu)$, eq.~(\ref{31}).

However, due to the very complicated structure of the excitation
energies, eq.~(\ref{213}), this is also a very difficult task. Even for
the determination of the heat-kernel coefficients, eq.~(\ref{32}), it is
not clear how to do it for the coefficients $b_{\pm}$ separately. To
explain the difficulties, let us write $\lambda_j$ for the eigenvalues
of the Laplacian, then we have
\beq
E_{\pm ,j}^2&=&\lambda_j +\mu^2+\frac 1 2 \massee\label{36}\\
&\pm&\left[2\mu^2\left\{2\lambda_j+\massee\right\}
+\frac 1 4 \massez\right]^{\frac 1 2}.\nn
\eeq
Here the difference to the case of vanishing background charge is seen
explicitly. In contrast to a second order elliptic differential operator
(to which eq.~(\ref{36}) reduces for $\mu =0$) in this case one has a
pseudo-differential operator. Whereas in the first case one could use
very powerful results of mathematicians and physisists for this kind of
operators, to our knowledge no results on the heat-kernel expansion,
eq.~(\ref{32}), are available.
However, as we will see in the actually needed sum of
$\Omega_+$ and $\Omega_-$ we avoid the mentioned problem and the
calculation
of the coefficient $b_l^{\pm}$ reduces to the known case $\mu =0$.

For notational convenience let us write
\beq
d_j&=&\lambda_j+\frac 1 2 \massee,\nn\\
b &=&\frac 1 4 \massez,\nn
\eeq
thus the eigenvalues read
\beq
E_{\pm ,j}=d_j+\mu^2\pm \sqrt{4\mu^2 d_j +b}.\label{37}
\eeq
As we see in eq.~(\ref{212}), we need the asymptotic $t\to 0$ expansion
of
\beq
K(t)&=&K_+(t)+K_-(t)\nn\\
   &\sim&\left(\frac 1 {4\pi t}\right)^{\frac 3 2 }\sum_{l=0,\frac 1 2 ,
1,...}^{\infty}(b_l^++b_l^-)t^l.\label{38}
\eeq
Expansion of the exponential factor containing the square root leads to
the cancellation of terms containing explicitly a square root, letting
us with \beq
K(t) &\sim&2\sum_j\slnu e^{-(d_j+\mu^2)t}\frac{t^{2l}}{(2l)!}
(4\mu^2d_j+b)^l.\label{39}
\eeq
The powers in $d_j$ may be produced by differentiation of the
exponential, which then yields a reduction to the heat-kernel connected
with $d_j$ which is nowadays a very well known problem. In doing that,
it is seen, that the leading term for $t\to 0$ of the summation index
$l$ is given by $t^{-\frac 3 2 +l}$. Thus, reducing our attention to the
coefficients $b_0,...,b_2$, $b_l=b_l^+ +b_l^-$, we find that the
following terms may contribute
\beq
K(t)=A_0+A_1+A_2+\cao (t)\label{310}
\eeq
with
\beq
A_0 &=&2\left(1-\mu^2t+\frac 1 2 \left[b+\mu^4\right]t^2+\cao
(t^3)\right)\sum_j e^{-d_j t},\nn\\
A_1&=&4\mu^2 t^2\left(1-\mu^2 t+\cao (t^2)\right)\sum_j
d_j e^{-d_jt},\label{311}\\
A_2&=&\frac 4 3 \mu^4t^4 (1+\cao(t))\sum_j d_j^2e^{-d_j t}.\nn
\eeq
Introducing the kernel of $d_j$,
\beq
B(t)=\sum_j e^{-a_j t}\sim\left(\frac 1 {4\pi t}\right)^{\frac 3 2}
    \sum_{l=0,\frac 1 2, 1,...}c_lt^l,\label{312}
\eeq
this may be rewritten as
\beq
A_0&=&
\tdh \times 2\left(c_0+c_{\frac 1 2}t^{\frac 1 2}+[c_1-c_0\mu^2]t+
[c_{\frac 3 2}-\mu^2c_{\frac 1 2}]t^{\frac 3 2}\right.\nn\\
& &\left.+[c_2-\mu^2 c_1+\frac 1 2 (b+\mu^4)c_0]t^2+\cao (t^{\frac 5
2})\right),\nn\\
A_1&=&\tdh\times\left(6c_0\mu^2 t+4\mu^2c_{\frac 1 2}t^{\frac 3 2}
+2[c_1\mu^2-3c_0\mu^4]t^2+\cao(t^{\frac 5
2})\right),\hspace{2mm}  \phantom{12}  \label{313}\\
A_2&=&\tdh\times \left(5\mu^4c_0t^2+\cao (t^{\frac 5 2})\right).\nn
\eeq
Using the knowledge of the heat-kernel coefficients $c_j$, this solves
already the task of finding the high-temperature behaviour of the
considered theory and it can be done in principle to any desired order.
Summing up the contributions in equation (\ref{313}) we obtain to the
considered order
\beq
K(t)&\sim& \left(\frac 1 {4\pi t}\right)^{\frac 3 2}\left\{
2c_0+2c_{\frac 1 2}t^{\frac 1
2}+2\left[c_1+2c_0\mu^2\right]t\right.\label{314}\\
& &\phantom{aaa}\left.  +2\left[c_{\frac 3 2}+c_{\frac 1
2}\mu^2\right]t^{\frac 3 2}+2\left[c_2+\frac 1 2 bc_0\right]t^2+\cao
\left(t^{\frac 5 2}\right)\right\}.\nn
\eeq
Thus the leading terms of the thermodynamic potential read
\beq
\Omega&\sim& -\frac{\pi^2}{45}c_0T^4-\frac 1 {2\pi^{\frac 3
2}}c_{\frac 1 2}\zeta_R(3)T^3\label{315}\\
& &-\frac 1 {12}\left[c_1+2c_0\mu^2\right]T^2+\cao (T)\nn
\eeq
For a manifold without boundary this is for the effective potential,
\beq
V_{thermal}&\sim&-\frac{\pi^2}{45}T^4-\frac 1 6 \mu^2T^2+\frac 1 {12}
M^2T^2+\cao (T)\label{316}
\eeq
with
\beq
M^2=\frac 1 2 \massee -\frac 1 6 R.\nn
\eeq
For the interaction potential
\beq
U(\phi)=\frac 1 2 (m^2+\xi R)\phi^2+\frac{\lambda}{4!}\phi^4\nn
\eeq
we have
\beq
M^2=m^2+\q R+\frac 1 3 \lambda \cl^2\nn
\eeq
generalizing the flat space result \cite{haber,benson,bernstein} to
curved spacetimes.

\subsection{Weak coupling expansion}
Our second approach is based on an expansion in the self-interaction
potential. In doing that, let us also present another very simple way to
determine the excitation energies $E_{\pm}$. For convenience we
introduce here a comlex field $\Phi =\frac 1
{\sqrt{2}}(\Phi_1+i\Phi_2)$.
Then, the action (\ref{eq:1.11}) is given by
\beq
\bar S &=&\int           dt \ins \left\{(\dot{\Phi}^{\dagger}
+i\mu\Phi^{\dagger})
(\dot{\Phi} -i\mu\Phi) -|\nabla\Phi|^2\right.\nn\\
& &\left.-(m^2 +\xi R)|\Phi ^2| -V_{tree}(\Phi )\right\}.
\label{eq:4.9}
\eeq
The field equations for $\Phi$ and $\Phi^\dagger$
follow from $\bar{S}$ in the
usual
way by varying with respect to $\Phi$ and $\Phi^\dagger$ independently.
It is easily seen that
\begin{eqnarray}
          0&=&-\ddot{\Phi}+2i\mu\dot{\Phi}+\mu^2\Phi+\nabla^2\Phi-
               (m^2+\xi R)\Phi-V_{tree}'\Phi,\label{eq:4.10}\\
          0&=&-\ddot{\Phi}^\dagger+2i\mu\dot{\Phi}^\dagger+\mu^2
               \Phi^\dagger+\nabla^2\Phi^\dagger-
               (m^2+\xi R)\Phi-V_{tree}'\Phi^\dagger.\label{eq:4.11}
\end{eqnarray}
(Here $V_{tree}'=\frac{\partial}{\partial|\Phi|^2}
V_{tree}(|\Phi|^2)$.)
The excitation
energies are obtained by looking at small fluctuations about
$\bar{\Phi}$. We will write
\begin{equation}
          \Phi=\bar{\Phi}+\Psi\label{eq:4.12}
\end{equation}
and linearize (\ref{eq:4.10},\ref{eq:4.11}) in $\Psi$. This gives
\begin{eqnarray}
          0&=&-\ddot{\Psi}+2i\mu\dot{\Psi}+\mu^2\Psi+\nabla^2\Psi-
               (m^2+\xi R)\Psi\nonumber\\
          &&\quad-V_{tree}'(|\bar{\Phi}|^2)\Psi-V_{tree}''
(|\bar{\Phi}|^2)
             \Big\lbrack|\bar{\Phi}|^2\Psi+\bar{\Phi}^2
\Psi^\dagger\Big\rbrack
             \label{eq:4.13}
\end{eqnarray}
along  with  the  equation  obtained  by  taking  the  complex
conjugate  of
(\ref{eq:4.13}). It is of course not possible to solve
(\ref{eq:4.13}); however,
we
can expand $\Psi$ and $\Psi^\dagger$ in terms of a complete set of
functions.
Let  $\lbrace  f_N({\bf  x})\rbrace$  be  a  complete set of
eigenfunctions
 of
$-\nabla^2+\xi R$ with eigenvalues $\sigma_N$~:
\begin{equation}
          (-\nabla^2+\xi R)f_N({\bf x})=\sigma_Nf_N({\bf x})\;.
\label{eq:4.14} \end{equation}
We can write
\begin{eqnarray}
          \Psi(t,{\bf  x})&=&\sum_N A_Ne^{-iE_Nt}f_N({\bf
x}),\label{eq:4.15}\\
          \Psi^\dagger(t,{\bf     x})&=&
\sum_NB_Ne^{-iE_Nt}f_N({\bf     x}),
          \label{eq:4.16}
\end{eqnarray}
for some independent expansion coefficients $A_N$ and $B_N$.
Substitution
of (\ref{eq:4.15},\ref{eq:4.16}) into (\ref{eq:4.13}) and its complex
conjugate
 leads
to the coupled equations
\begin{eqnarray}
          0&=&\Big\lbrack(E_N+\mu)^2-\sigma_N-m^2-
V_{tree}'-V_{tree}''
             |\bar{\Phi}|^2\Big\rbrack A_N\nonumber\\
                &&\quad\quad\quad-\bar{\Phi}^2V_{tree}''B_N,
               \label{eq:4.17}\\
          0&=&\Big\lbrack(E_N-\mu)^2-
\sigma_N-m^2-V_{tree}'-V_{tree}''
             |\bar{\Phi}|^2\Big\rbrack B_N\nonumber\\
               &&\quad\quad\quad-(\bar{\Phi}^\dagger)^2V_{tree}''A_N.
               \label{eq:4.18}
\end{eqnarray}
In order for a non-trivial solution for $A_N$ and $B_N$ we must have
$E_N= E_{\pm N}$ where
\begin{equation}
          E_{\pm N}=\Big\lbrace \mu^2+M_N^2\pm\Big\lbrack
           4\mu^2M_N^2+|\bar{\Phi}|^4
\Big(V_{tree}''\Big)^2\Big\rbrack^{1/2}
            \Big\rbrace^{1/2}\;,\label{eq:4.19}
\end{equation}
with
\begin{equation}
 M_N^2=\sigma_N+m^2+V_{tree}'+|\bar{\Phi}|^2V_{tree}''\;.
\label{eq:4.20} \end{equation}
This gives the excitation energies of
Refs.~\cite{benson,bernstein} if the result is
 rewritten
in terms of real fields and it is the result (\ref{213})

Once the excitation energies have been determined, the thermodynamic
potential
follows from the usual definition. Alternatively we can deal with the
effective
potential or effective action. The thermodynamic potential $\Omega$ is
given by \begin{equation}
          \Omega=T\sum_N\ln\left[\Big(1-e^{-\beta
E_{+N}}\Big)\Big(1-e^{-\beta
          E_{-N}}\Big)\right]\;.\label{eq:5.1}
\end{equation}
(If we deal with the effective potential then there is a contribution
from the
zero-point energy \cite{benson}, but it can be ignored at high
temperature.) If
we set $V_{tree}=0$ so that the theory is free, then it is easy to see
from (\ref{eq:4.19}) and (\ref{eq:4.20}) that
\begin{equation}
          E_{\pm N}=(\sigma_N+m^2)^{1/2}\pm \mu\;.\label{eq:5.2}
\end{equation}
The term $(\sigma_N+m^2)^{1/2}$ just gives the energy of a single mode
$f_
N({\bf x})$, and is the analogue of the single particle energy $({\bf
k}^2+
m^2)^{1/2}$ in flat spacetime. Using (\ref{eq:5.2}) in (\ref{eq:5.1})
results in
the standard expression for the thermodynamic potential for a free
boson gas at finite density.

When $V_{tree}\ne0$, the evaluation of (\ref{eq:5.1}) becomes more
complicated.
A  general  approach  to  this  problem  can  be  made  using  generalized
$\zeta$-functions, which has been presented in the first part of this
section. However, it is very simple to obtain the leading term in
$\Omega$ coming from $V_{tree}$ at high temperature. To do this define
\begin{equation}
          M_N^2=\omega_N^2+\delta M^2\label{eq:5.3}
\end{equation}
where
\begin{eqnarray}
          \omega_N&=&(\sigma_N+m^2)^{1/2}\;,\label{eq:5.4}\\
          \delta M^2&=&V_{tree}'(|\bar{\Phi}|^2)+|\bar{\Phi}|^2V_{tree}''
            (|\bar{\Phi}|^2)\;.\label{eq:5.5}
\end{eqnarray}
If we work to first order in the potential it is easy to show that
\begin{equation}
          E_{\pm  N}=\omega_N\pm   \mu+\frac{1}{2\omega_N} \delta
M^2+
          \cdots\;.\label{eq:5.6}
\end{equation}
Using this in (\ref{eq:5.1}) and expanding to first order in $\delta
M^2$
 results
in
\begin{eqnarray}
          \Omega&=&T\sum_N\ln\Big(1-e^{-\beta(\omega_N+\mu)}\Big)
              \Big(1-e^{-\beta(\omega_N-\mu)}\Big)\nonumber\\
          &&\quad+\sum_N\frac{1} {2\omega_N}\delta
M^2\Big\lbrace\Big\lbrack
          e^{\beta(\omega_N+\mu)}-1\Big\rbrack^{-1}\nonumber\\
            &&\quad\quad+\Big\lbrack
          e^{\beta(\omega_N+\mu)}-
1\Big\rbrack^{-1}\Big\rbrace+\cdots\;.
           \label{eq:5.7}
\end{eqnarray}
The first term on the right hand side of (\ref{eq:5.7}) is just the
 thermodynamic
potential in the absence of interactions, $\Omega_{free}$,
\begin{equation}
          \Omega_{free}=T\sum_N\ln\Big(1-e^{-\beta(\omega_N+\mu)}\Big)
              \Big(1-e^{-\beta(\omega_N-\mu)}\Big)\;.\label{eq:5.8}
\end{equation}
To evaluate the second term on the right hand side of (\ref{eq:5.7}),
note that \begin{equation}
          \frac{\partial}{\partial m^2}\ln
\Big(1-e^{-\beta(\omega_N\pm \mu)}
              \Big) =\frac{\beta}{2\omega_N}\Big\lbrack
e^{\beta(\omega_N\pm
              \mu)}-1\Big\rbrack^{-1}\;.\label{eq:5.9}
\end{equation}
This shows that
\begin{eqnarray}
          \Omega_1&=&
          \sum_N\frac{1}{2\omega_N}\delta M^2\Big\lbrace\Big\lbrack
          e^{\beta(\omega_N+\mu)}-1\Big\rbrack^{-1}+\Big\lbrack
          e^{\beta(\omega_N+\mu)}-1\Big\rbrack^{-1}
\Big\rbrace\nonumber\\
          &=&\delta M^2\frac{\partial}{\partial m^2}\Omega_{free}\;.
           \label{eq:5.10}
\end{eqnarray}
The first order correction to the thermodynamic potential due to the
interaction can be determined from a knowledge of $\Omega_{free}$.

Expressions  for  $\Omega_{free}$ can  be  found  in  a  number  of
places
\cite{heat,bose,doktor}. The result for a manifold with boundary is (for
$D=3$) \begin{eqnarray}
          \Omega_{free}&=&-\frac{\pi^2}{45}T^4\theta_0-\frac{1}{2}
                     \pi^{-3/2}\zeta_R(3)T^3\theta_{1/2}\nonumber\\
          &&\quad-\frac{1}{12}T^2\Big\lbrack\theta_1-(m^2-2\mu^2)
              \theta_0\Big\rbrack+\cdots\label{eq:5.11}
\end{eqnarray}
at high temperature. $\theta_k$ are the heat kernel coefficients for
${\rm tr}\exp\Big(-t(-\nabla^2+\xi R)\Big)$. The result for spatial
dimensions other than
3 is also easily obtained~:
\begin{eqnarray}
          \Omega_{free}&=&-2\pi^{-(D+1)/2}\Gamma\Big((D+1)/2\Big)
           \zeta_R(D+1)T^{D+1}\theta_0\nonumber\\
          &&-\pi^{-(D+1)/2}\Gamma(D/2)\zeta_R(D)
T^D\theta_{1/2}\nonumber\\
          &&-\frac{1}{2}\pi^{-(D+1)/2}\Gamma
\Big((D-1)/2\Big)\zeta_R(D-1)
           T^{D-1}\nonumber\\
         &&\quad\times\Big\lbrace\theta_1-\Big\lbrack
m^2-(D-1)\mu^2
        \Big\rbrack\theta_0\Big\rbrace+\cdots\label{eq:5.12}
\end{eqnarray}
for $D\ge3$.

If we consider only $D=3$, and use $\theta_0=V$, then
\begin{equation}
          \frac{\partial}{\partial m^2}\Omega_{free}
=\frac{1}{12}T^2V+\cdots
          \label{eq:5.13}
\end{equation}
and so from (\ref{eq:5.10})
\begin{equation}
          \Omega_1=\frac{1}{12}T^2V\delta M^2+\cdots\;.\label{eq:5.14}
\end{equation}
We therefore have found the high temperature expansion
\begin{eqnarray}
          \Omega&=&-\frac{\pi^2}{45}VT^4-\frac{1}{2}\pi^{-3/2}
\zeta_R(3)
            T^3\theta_{1/2}\nonumber\\
          &&+\frac{1}{12}T^2\Big\lbrack(\delta M^2+m^2-2\mu^2)
             V-\theta_1\Big\rbrack+\cdots\label{eq:5.15}
\end{eqnarray}
$\delta M^2$ is given directly in terms of the interaction potential
by (\ref{eq:5.5}). If
we specialize to $\Sigma$ having no boundary then $\theta_{1/2}=0$
and $\theta_1=-(\xi-1/6)RV$. We therefore find
\begin{eqnarray}
          V_{thermal}&=&\Omega/V\nonumber\\
          &=&-\frac{\pi^2}{45}T^4+\frac{1}{12}T^2\Big\lbrack V_{tree}'
 (|\bar{\Phi}|^2)+|\bar{\Phi}|^2V_{tree}''(|\bar{\Phi}|^2)\nonumber\\
          &&+m^2+(\xi-1/6)R-2\mu^2\Big\rbrack+\cdots\;.
\label{eq:5.16} \end{eqnarray}
in agreement with eq.~(\ref{316}).
It is possible to calculate higher order terms in the interaction
potential by
obtaining the second order correction to the excitation energies.
However it is
clear on dimensional grounds that such terms will only involve lower
powers
of $T$, and therefore that these terms will be negligible at high
temperature.
\section{$O(N)$ model at finite chemical potential in the limit $N \to
\infty$}
\setcounter{equation}{0}
Let us conclude the considerations on the $O(N)$ model with two comments
on the limit $N\to \infty$.

The first comment is on technical simplifications that occur.
In the large-$N$ limit the
off diagonal elements of the potential matrix $V(\cl )$,
eq.~(\ref{eq:1.16}), may be neglected \cite{haber}, so that the
potential reduces to
\beq
V_{ij} =\delta_{ij} \left(\frac{V'_{tree} (\cl
)}{\cl}\right).\label{eq:6.1}
\eeq
Thus the fluctuation operator $\bar S _{ij}[\cl ]$ splits into $N/2$
$2\times 2$ matrices,
\beq
\bar S_{\alpha} [\cl ]&=&
\left(\begin{array}{cc}
  \bar S _{2\alpha -1,2\alpha -1}    &     \bar S _{2\alpha
-1,2\alpha}\\
     \bar S _{2\alpha,2\alpha -1}    &     \bar S _{2\alpha ,2\alpha}
      \end{array}\right) \nn\\  &=&
\left(\begin{array}{cc}
     -\Box -\mu_{\alpha}^2+M_2^2    &     2i \mu_{\alpha}
                               \frac{\partial}{\partial\tau}\\
     - 2i \mu_{\alpha}\frac{\partial}{\partial\tau}  &
                           -\Box -\mu_{\alpha}^2+M_2^2
      \end{array}\right).  \label{eq:6.2}
\eeq
So it is easily seen, that also the one-loop contribution splits into
the sum of $N/2$ contributions,
\beq
\ln \caz ^{(1)}=-\frac 1 2 \sum_{\alpha}^{N/2}\ln\det\bar S _{\alpha}
 [\cl ].\label{eq:6.3}
\eeq
Thus in the limit of large $N$, the one-loop contribution is simply the
sum of the contribution of $N/2$ complex fields with their associated
chemical potentials.

The second comment is on the influence of higher-loop contributions in
the limit $N\to \infty$.
As is well known, at the critical temperature the one-loop approximation
is not valid any more. This is made obvious by the fact, that the exact
one-loop contribution would lead to an unacceptable complex critical
temperature \cite{dolanjackiw}. The reason is that higher loop
contribution become significant and all so called daisy diagrams have to
be summed \cite{dolanjackiw,fendley}. Let us look in some detail at
this summation for the $O(N)$ theory in curved spacetime and let us
restrict for definiteness on the $\lambda \Phi^4$ self-interaction. Then
in principle, the expansion around the background field leads also to
odd powers in $\Phi$, but in the limit $N\to \infty$ these are
negligible \cite{dolanjackiw}. The relevant quantity to tackle with the
problem is the 1PI self-energy, which one may approximate by
\cite{fendley}
\beq
\sigma \sim \frac{\lambda TN}6 \sun \sum_l \frac 1
{(2\pi/\beta)^2n^2+\lambda_l},\label{eq:7.1}
\eeq
with the eigenvalues $\lambda_l$ of the Laplace-Beltrami operator
$\Delta _{\Sigma}$. Let us regularize the expression by considering
\beq
\zeta_{\sigma}(s) =\frac{\lambda TN} 6 \sun \sum_l
   \left(    \left(\frac{2\pi}{\beta}\right)^2n^2+\lambda_l \right)^{-s}
   \label{eq:7.2}
\eeq
and lets derive a high-temperature expansion of that expression. Using
the method described in section III, one arrives at
\beq
\zeta_{\sigma} (s) &=& \frac{\lambda TN} 6 \zeta_{\Delta} (s)\nn\\
 & &+\frac {\lambda TN}{3(4\pi)^{\frac 3 2 }\Gamma(s) } \sum_{l=0}
   ^{\infty} \Gamma \left(s-\frac 3 2 +l\right) a_l (\Delta )
        (2\pi T)^{3-2s-2l}\zeta_R (2s-3+2l ),\label{eq:7.3}
\eeq
with the zeta function $\zeta_{\Delta}$ and heat-kernel coefficients
$a_l (\Delta )$ of the operator $\Delta_{\Sigma}$. From general zeta
function theory it follows, that depending on spacetime properties there
might exist a pole at $s=1$. This divergent piece has to be dealed with
in the zero temperature renormalization theory. The leading terms of the
finite part of the self-energy $\sigma$ defined by the finite part $PP$
of $\zeta_{\sigma} (s)$, are
\beq
\sigma \sim \frac{\lambda N T^2}{72} +\frac{\lambda NT} 6
PP\zeta_{\Delta} (1). \label{eq:7.4}
\eeq
The leading term is exactly the flat spacetime result, as to be expected
for dimensional reasons, and so also in curved spacetime the leading
orders of the expansion are not influenced by summing over all daisy
diagrams.

\section{Extension to general gauge groups}
\setcounter{equation}{0}

We have, following Haber and Weldon \cite{haber}, studied the $O(N)$
model for
$N$
even. In this section we wish to consider the more general case of an
arbitrary
gauge group $G$ with a set of interacting complex scalar fields with a
rigid gauge invariance under the action of $G$.

Let $\Phi$ denote the set of complex scalar fields transforming under
some $d_s$-dimensional representation of $G$. We can write
\begin{equation}
        \Phi\rightarrow U\Phi\label{eq:8.1}
\end{equation}
where
\begin{equation}
        U=\exp(i\theta^iT_i)\label{eq:8.2}
\end{equation}
with $T_i$ the hermitian generators for the Lie algebra of $G$ taken
in the
$d_s$-dimensional representation for the scalar fields. We will use
$i=1,2,\ldots,d_G$ where $d_G  $ is the dimension of $G$, and
$\alpha=1,2,\ldots,d_s$. Because we choose to work with complex fields
rather than real fields here, the action is
\begin{equation}
        S=\int dt\int_{\Sigma}d\sigma_x\Big\lbrace(\partial^\mu
\Phi^\dagger)
           (\partial_\mu\Phi)-m^2\Phi^\dagger\Phi-\xi
R\Phi^\dagger\Phi-V_{tree}\Big\rbrace\label{eq:8.3}
\end{equation}
where $\Phi$ may be regarded as a column vector of complex scalar
fields with
$d_s$ components. $V_{tree}$ is assumed to be invariant under
(\ref{eq:8.1}),
and for simplicity we will assume that it only depends on $\Phi$ and
$\Phi^\dagger$ through the combination $|\Phi|^2=\Phi^\dagger\Phi$ as
before.

A set of conserved currents can be found in the usual way (see
Ref.~\cite{Abers} for example) by looking at the change in
(\ref{eq:8.3}) under
an infinitesimal gauge transformation. It is easy to show that
\begin{equation}
        J_j^\mu=i\Big(\Phi^\dagger T_j\partial^\mu\Phi-
\partial^\mu\Phi^\dagger
           T_j\Phi\Big)\label{eq:8.4}
\end{equation}
gives the conserved currents, and hence
\begin{equation}
        Q_j=i\int_{\Sigma}d\sigma_x\Big(\Phi^\dagger
T_j\dot{\Phi}-\dot{\Phi}^\dagger T_j\Phi\Big)\label{eq:8.5}
\end{equation}
gives a set of conserved charges. For the case of $O(N)$ these results
reduce to those of Haber and Weldon \cite{haber} given in Sec.~2.

A chemical potential $\mu^i$ can be introduced for each charge $Q_i$,
and a
Hamiltonian $\bar{H}$ found which incorporates the conserved charges~:
\begin{equation}
        \bar{H}=H-\mu^iQ_i\;.\label{eq:8.6}
\end{equation}
It is now possible to integrate out the field momenta in the
Hamiltonian
path integral as before to leave a path integral in configuration
space. This results in
\begin{eqnarray}
        \bar{S}&=&\int dt\int_{\Sigma}d\sigma_x\Big\lbrace\Big(
        \dot{\Phi}^\dagger+i\mu^j\Phi^\dagger T_j\Big)
\Big(\dot{\Phi}-i\mu^j
          T_j\Phi\Big)-|\nabla\Phi|^2\nn\\
        &&\quad-(m^2+\xi R)|\Phi|^2-V_{tree}\Big\rbrace\;,
\label{eq:8.7} \end{eqnarray}
as the replacement for (\ref{eq:4.9}). (If we specialize to $O(N)$,
decompose
the fields into real and imaginary parts, and perform a Wick rotation to
imaginary time, then the earlier result in (\ref{eq:1.11}) is found.)

In order to calculate the thermodynamic potential we will content
ourselves
with a high temperature expansion. This can be done in two stages. We
will
first look at the case where the chemical potentials all vanish as we
did in
Sec.~3 for the $O(N)$ model. However we will use the technique based
on the excitation energies used in Sec.~4.2 for the $U(1)$ model. After
showing how
the thermodynamic potential can be obtained in this case, we will then
return
to the case of non-vanishing chemical potentials and show how a
perturbative
expansion may be used to obtain the leading terms of the thermodynamic
potential in the high temperature limit.

\subsection{Vanishing chemical potentials}

Setting $\mu^i=0$ in (\ref{eq:8.7}) and varying with respect to $\Phi$
and $\Phi^\dagger$
results in the field equation
\begin{equation}
          0=-\ddot{\Phi}+\nabla^2\Phi-(m^2+\xi R)\Phi
 -V_{tree}'\Phi\;,\label{eq:8.8}
\end{equation}
and its complex conjugate. As in Sec.~4.2, let $\Phi=\bar{\Phi}+\Psi$
and
 $\Phi^\dagger=\bar{\Phi}^\dagger
+\Psi^\dagger$ where $\bar{\Phi}$ denotes the background field.
Linearizing in
 $\Psi$
results in
\begin{eqnarray}
          0&=&-\ddot{\Psi}+\nabla^2\Psi-(m^2+\xi
 R)\Psi-V_{tree}'(|\bar{\Phi}|^2)\Psi\nn\\
 &&\quad-V_{tree}''(|\bar{\Phi}|^2)(\bar{\Phi}^\dagger\Psi+
\Psi^\dagger\bar{\Phi
 })\bar{\Phi}\;,\label{eq:8.9}
\end{eqnarray}
and its complex conjugate. By using expansions as in (\ref{eq:4.15}) and
 (\ref{eq:4.16}),
except that $A_N$ is a column vector and $B_N$ is a row vector, we
obtain the coupled set of equations
\begin{eqnarray}
          0&=&\lbrack(E_N^2-m^2-\sigma_N-V_{tree}')\delta_{\alpha\beta}
               -V_{tree}''\bar{\phi}^\dagger_{\beta}\bar{\Phi}_
\alpha\rbrack
                  A_{N\beta}\nn\\
          &&\quad-V_{tree}''\bar{\Phi}_\alpha\bar{\Phi}_\beta
 B_{N\beta}\;,\label{eq:8.10a}\\
          0&=&  -V_{tree}''\bar{\phi}^\dagger_{\alpha}
                  \bar{\Phi}_\beta^\dagger
                    A_{N\beta}+B_{N\beta}
               \lbrack(E_N^2-m^2-\sigma_N-V_{tree}')
\delta_{\beta\alpha}\nn\\
         &&\quad  -V_{tree}''\bar{\phi}_{\beta}
\bar{\Phi}_\alpha^\dagger\rbrack
              \;.\label{eq:8.10b}
\end{eqnarray}
One simple way of obtaining the energy spectrum is by decomposing
\begin{equation}
          A_{N\beta}=A_{N\beta}^{\perp}+\bar{\Phi}_\beta
            A_{N}^{\parallel}\label{eq:8.11}
\end{equation}
where
\begin{equation}
          \bar{\Phi}^{\dagger}_\beta
A_{N\beta}^{\perp}=0\label{eq:8.12} \end{equation}
and similarly
\begin{equation}
          B_{N\beta}=B_{N\beta}^{\perp}+\bar{\Phi}_\beta^\dagger
               B_{N}^{\parallel}\label{eq:8.13}
\end{equation}
where
\begin{equation}
          \bar{\Phi}_\beta B_{N\beta}^{\perp}=0\;.\label{eq:8.14}
\end{equation}
Using these decompositions in (\ref{eq:8.10a}) and (\ref{eq:8.10b})
results in \begin{eqnarray}
          0&=&aA_{N\alpha}^{\perp}+\bar{\Phi}_\alpha
\lbrace(a-|\bar{\Phi}|^2
                 V_{tree}'')A_N^{\parallel}-|\bar{\Phi}|^2 V_{tree}''
                   B_N^{\parallel}\rbrace\label{eq:8.15}\\
 0&=&aB_{N\alpha}^{\perp}+\bar{\Phi}_\alpha^\dagger
\lbrace(a-|\bar{\Phi}|^2
                 V_{tree}'')B_N^{\parallel}- |\bar{\Phi}|^2 V_{tree}''
                   A_N^{\parallel}\rbrace\label{eq:8.16}
\end{eqnarray}
where
\begin{equation}
          a=E_N^2-m^2-\sigma_N-V_{tree}'\;.\label{eq:8.17}
\end{equation}
By contracting (\ref{eq:8.15}) with $\bar{\Phi}_\alpha^\dagger$ and
 (\ref{eq:8.16})
with $\bar{\Phi}_\alpha$ it is easy to show that $a=0$ or $a=
 2|\bar{\Phi}|^2V_{tree}''$,
where $a=0$ occurs with multiplicity $d_s-1$ and $a=
2|\bar{\Phi}|^2V_{tree}''$
 occurs
with multiplicity 1. The excitation energies are therefore
\begin{eqnarray}
          E_{+N}&=&\lbrack m^2+\sigma_N+V_{tree}'+ 2|\bar{\Phi}|^2
                  V_{tree}''\rbrack^{1/2}\;,\label{eq:8.18}\\
          E_{-N}&=&\lbrack m^2+\sigma_N+V_{tree}'
\rbrack^{1/2}\;.\label{eq:8.19} \end{eqnarray}
These results agree with the $\mu\rightarrow0$ limit of
(\ref{eq:4.19}) and (\ref{eq:4.20}).

We therefore find that the thermodynamic potential for $\mu=0$ is
\begin{equation}
          \Omega(\mu=0)=T\sum_{N}\Big\lbrace  \ln\big(1-e^{-\beta
E_{+N}}
                \big)+(d_s-1) \lbrace\ln\big(1-e^{-\beta
 E_{-N}}\big)\big\rbrace\;.\label{eq:8.20}
\end{equation}
It is now possible to make contact with the $\zeta$-function method of
Secs. 2,3
to obtain the high temperature expansion. Instead of pursuing this we will
 simply
obtain the leading order terms in the high temperature expansion using the
 weak-coupling
limit described in Sec. 4.2. It is straightforward to show that
\begin{eqnarray}
          \Omega(\mu=0)&\simeq&T\sum_N\big \lbrace
d_s\ln(1-e^{-\beta\omega_N})
 +\frac{\beta}{2\omega_N}(e^{\beta\omega_N}-1)^{-1}\nn\\
          &&\quad\times\lbrack
 d_sV_{tree}'+2|\bar{\Phi}|^2V_{tree}''\rbrack\big\rbrace
\label{eq:8.21}\\
          &=&\frac{1}{2}\big\lbrace d_s+\lbrack
 d_sV_{tree}'+2|\bar{\Phi}|^2V_{tree}''\rbrack
              \frac{\partial}{\partial
 m^2}\big\rbrace\Omega_{free}(\mu=0)\label{eq:8.22}
\end{eqnarray}
where
\begin{equation}
          \Omega_{free}(\mu=0)=T\sum_N\ln(1-e^{-\beta\omega_N})^2
\label{eq:8.23} \end{equation}
is the thermodynamic potential for a single complex scalar field with no
 interactions
in the absence of a chemical potential. In the case of $D=3$, with no
boundary effects included
\begin{equation}
 \Omega_{free}(\mu=0)\simeq-\frac{\pi^2}{45}VT^4+\frac{1}{12}T^2V
\lbrack
                m^2+(\xi-\frac{1}{6})R\rbrack+\cdots\;.\label{eq:8.24}
\end{equation}
(It is possible to generalize this to include boundary effects and
spacetimes
whose
dimension differs from 4 by using the results quoted in Sec. 4.2.)

We therefore find
\begin{eqnarray}
          \Omega(\mu=0)&\simeq&-\frac{\pi^2}{90}d_sVT^4+\frac{1}{24}T^2V
            \lbrace d_s\lbrack m^2+(\xi-\frac{1}{6})R\nn\\
 &&\quad+V_{tree}'\rbrack+2|\bar{\Phi}|^2V_{tree}''\rbrace+\cdots
\label{eq:8.25} \end{eqnarray}
as the leading terms in the high temperature expansion. For the $U(1)$
model,
which
corresponds to $d_s=2$ (although there is only one complex scalar
field the
dimension
refers to the real dimension), we recover the $\mu\rightarrow0$ result
of
(\ref{eq:5.16}).
For the $O(N)$ model it agrees with (\ref{eq:2.25}).

\subsection{Inclusion of the chemical potential}

In order to find the leading term at high temperature when the chemical
 potential
is non-zero we will treat the terms in the action (\ref{eq:8.7}) which
involve
 the
chemical potential as an interaction and work to quadratic order in
the chemical
potential. It is easiest to use the imaginary time formalism in which
 (\ref{eq:8.7})
is replaced by
\begin{eqnarray}
          \bar{S}&=&\int_{0}^{\beta}dt\int_\Sigma
 d\sigma_x\lbrace(\dot{\Phi}^\dagger+
             \mu^j\Phi^\dagger T_j)(\dot{\Phi}-\mu^j
 T_j\Phi)+|\nabla\Phi|^2\nn\\
          &&\quad+(m^2+\xi R)|\Phi|^2+V_{tree}\rbrace\label{eq:8.26}
\end{eqnarray}
after the Wick rotation to imaginary time. We will define
\begin{equation}
          \bar{S}_{int}=\int_{0}^{\beta}dt\int_\Sigma d\sigma_x
\lbrace\mu^j
                (\Phi^\dagger T_j\dot{\Phi}-\dot{\Phi}^\dagger T_j\Phi)
                 -\mu^j\mu^k\Phi^\dagger T_jT_k\Phi\rbrace\label{eq:8.27}
\end{equation}
as the interaction part of the action. We are therefore perturbing
around the $\mu=0$
result for which we know the thermodynamic potential already from
Sec.~7.1.

Up to second order in the interaction we have
\begin{equation}
          \Gamma=\langle\bar{S}_{int}\rangle-\frac{1}{2}\langle
                \bar{S}_{int}^2\rangle\label{eq:8.28}
\end{equation}
where terms denoted by $\langle\cdots\rangle$ are to be Wick reduced
with only
one-particle irreducible graphs kept. (See Ref.~\cite{background} for
example.) The
basic result is
\begin{equation}
          \langle\Phi_\alpha(x)\Phi_\beta^\dagger(x')\rangle=
            G_{\alpha\beta}(x,x')\label{eq:8.29}
\end{equation}
where
\begin{equation}
          \left(-\frac{\partial^2}{\partial t^2}-\nabla^2+m^2+ \xi
R\right)
 G_{\alpha\beta}(x,x')=\delta_{\alpha\beta}\delta(x,x')\label{eq:8.30}
\end{equation}
defines the thermal Green function. (We have dropped the terms in
$V_{tree}$
 here
as the first order correction due to the self-interaction was
considered
earlier.)

We have
\begin{equation}
          \langle\bar{S}_{int}\rangle=-\mu^j\mu^k(T_jT_k)_{\alpha\beta}
        \int_{0}^{\beta}dt\int_\Sigma
 d\sigma_x\,G_{\alpha\beta}(x,x)\;.\label{eq:8.31}
\end{equation}
Because we are only after the leading behaviour at high temperature,
we may
approximate
$G_{\alpha\beta}(x,x')$ by its flat spacetime expression. We have
\begin{eqnarray}
          G_{\alpha\beta}(x,x)&\simeq&\delta_{\alpha\beta}
           \sum_{n=-\infty}^{\infty}\frac{1}{\beta}
          \int\frac{d^Dk}{(2\pi)^D}\left\lbrack\left(\frac{2\pi
 n}{\beta}\right)^2
              +k^2+m^2\right\rbrack^{-1}\nn\\
          &=&\delta_{\alpha\beta}T(4\pi)^{-D/2}(2\pi T)^{D-2}
\Gamma(1-D/2)\nn\\
          &&\quad\times\left\lbrace\left(\frac{m}{2\pi T}\right)^{D-2}
             +2\sum_{n=1}^{\infty}\left\lbrack n^2+
      \left(\frac{m}{2\pi
 T}\right)^2\right\rbrack^{D/2-1}\right\rbrace\;.\label{eq:8.32}
\end{eqnarray}
For high $T$, taking $D\rightarrow3$ it is easily seen from (\ref{eq:8.32})
that
\begin{equation}
 G_{\alpha\beta}(x,x)\simeq\frac{1}{12}T^2\delta_{\alpha\beta}\;.
\label{eq:8.33} \end{equation}
This leads to
\begin{equation}
          \langle\bar{S}_{int}\rangle\simeq-\frac{1}{12}TV
                 \mu^i\mu^j\,{\rm tr}(T_iT_j)\label{eq:8.34}
\end{equation}
as the leading contribution at high temperature.

The Wick reduction of the second term in (\ref{eq:8.28}), keeping only the part
 quadratic
in the chemical potential, results in
\begin{eqnarray}
          \langle\bar{S}_{int}^2\rangle&=&4\mu^i\mu^j
               (T_i)_{\alpha\beta}(T_j)_{\gamma\delta}
          \int_{0}^{\beta}dt\int_{0}^{\beta}dt'\int_\Sigma d\sigma_x
             \int_\Sigma d\sigma_{x'}\nn\\
          &&\quad\times\frac{\partial}{\partial t'}G_{\beta\gamma}(x,x')
            \frac{\partial}{\partial t}G_{\delta\alpha}(x',x)
\label{eq:8.35} \end{eqnarray}
after a short calculation. If we again use the flat spacetime Green
function,
with
the understanding that we will only keep the leading order term, it
may be shown from (\ref{eq:8.35}) that
\begin{eqnarray}
          \langle\bar{S}_{int}^2\rangle&=&-8\mu^i\mu^j{\rm tr}
             (T_iT_j)V(4\pi)^{-D/2}\Gamma(2-D/2)\nn\\
          &&\quad\times\sum_{n=1}^{\infty}(2\pi nT)^2
    \lbrack(2\pi nT)^2+m^2\rbrack^{D/2-2}\;.\label{eq:8.36}
\end{eqnarray}
Taking the high temperature limit and letting $D\rightarrow3$ results in
\begin{equation}
          \langle\bar{S}_{int}^2\rangle\simeq
     \frac{1}{6}TV\mu^i\mu^j{\rm tr}(T_iT_j)\;.\label{eq:8.37}
\end{equation}

Using (\ref{eq:8.34}) and (\ref{eq:8.37}) in (\ref{eq:8.28}) and
noting that $\Gamma=\beta\Omega$, we have
\begin{equation}
          \Omega\simeq-\frac{1}{6}T^2V\mu^i\mu^j{\rm tr}(T_iT_j)
\label{eq:8.38} \end{equation}
as the leading order correction to the thermodynamic potential at high
temperature
when the chemical potentials are non-zero.

It is possible to relate ${\rm tr} (T_iT_j)$ to the quadratic Casimir
invariant of the group using ${\rm tr} (T_iT_j) =(d_R C_2
(G_R)/N)\delta_{ij}$, where $d_R$ is the real dimension of the
representation $G_R$ of the gauge group, and $N$ is the dimension of the
group. In the special case of $O(N)$ the result in (\ref{eq:8.38})
reproduces (\ref{eq:2.25}) given earlier.

\section{Bose-Einstein condensation}
\setcounter{equation}{0}
Let us finally come to the application of the calculations to the
phenomenon of Bose-Einstein condensation for the case of a constant
background field. The high-temperature effective action $\Gamma
[\bar{\Phi}]$ including the classical part of the background field
$\bar{\Phi}$ has the form
\beq
\lefteqn{
\frac{\Gamma[\bar{\Phi}]}{\beta V} = (m^2 +\xi R) \clnorm -\mu ^i
\cl^{\dagger} T_i \mu ^j T_j\cl +V_{tree} (\cl )-\frac{\pi^2}{90} d_s
T^4} \label{7.1}\\
& &+\frac 1 {24} T^2 \left\{d_s \left[m^2 +\left(\xi -\frac 1 6 \right)
R +V'_{tree} (\cl ) \right] +2\clnorm V''_{tree} (\cl )\right\} -\frac 1
6 T^2 \mu ^i \mu ^j \tr (T_iT_j).\nn
\eeq
Thus we find the equations of motion
\beq
\left[m^2 +\xi R -\mu ^i T_i \mu^j T_j +V'_{tree} (\cl )+\frac{T^2}{24}
 \left[d_s V''_{tree} (\cl ) +2V''_{tree} (\cl
)\right]\right]\cl =0\label{7.2}
\eeq
together with its conjugate. Hence the effective action has one minimum
with unbroken symmetry, $\cl  =0$, and solution(s) with broken
symmetry determined by
\beq
\det \left(
m^2 +\xi R -\mu ^i T_i \mu^j T_j +V'_{tree} (\cl )+\frac{T^2}{24}
 \left[d_s V''_{tree} (\cl ) +2V''_{tree} (\cl )\right]\right)
=0.\label{7.3}
\eeq
For the $\cao (N)$-model introduced in section 2 this is equivalent to
\beq
\clnorm =\frac 3 {\lambda} \left[\mu^{\alpha} \mu_{\alpha} -\left( m^2
+\xi R +\frac{N+2}{72} \lambda T^2\right) \right].\label{7.4}
\eeq
For $N=2$ this generalizes the flat space result \cite{bernstein,benson}
and reduces to the result found in \cite{diag}.

The discussion of eq. (\ref{7.4}) then parallels completely the flat
space discussion \cite{haber} and we will indicate it only briefly for
$N=2$. At very high temperature one finds the charge density
\beq
\rho =-\frac 1 {\beta} \frac{\partial \Gamma}{\partial \mu} =+\mu
\clnorm +\frac 1 3 T^2 \mu +...,\label{7.5}
\eeq
the first piece corresponding to the charge $Q_0$ in the ground state,
the second one, $Q_1$, to the excited states. In the symmetric phase,
fixing the charge $Q$ and volume $V$, one thus have $\mu =3\rho / T^2$.
As temperature decreases, $\mu$ will increase and will reach a value at
which
\beq
\mu (T_c)=\frac{3\rho}{T_c^2} =\left( m^2+\xi R +\frac{\lambda}{18}
T_c^2\right) ^{\frac 1 2}.\label{7.6}
\eeq
Here $T_c$ is, corresponding to  (\ref{7.4}), the critical temperature
\beq
T_c^2 =\frac{18}{\lambda} \left[\mu^2 (T_c)-m^2 -\xi
R\right]\label{7.7} \eeq
at which Bose-Einstein condensation and thus symmetry breaking occurs.
For the charge in the ground state one then easily finds
\beq
Q_0 = Q\left[1-\left(\frac T {T_c}\right)^2\right],\label{7.8}
\eeq
which is identical to the form of the flat space result
\cite{haber,bernstein,benson}.
\section{Conclusions}
\setcounter{equation}{0}
In this paper we continued the analysis started in the references [1-5]
from flat to curved spacetime. Especially, we considered as a model a
set of interacting scalar fields with a rigid gauge invariance under the
action of an arbitrary gauge group $G$. In order to examine the
phenomenon of Bose-Einstein condensation we derived the high temperature
behaviour of the effective action of the theory. To this aim, different
approaches were developed, one being based on the zeta function approach
in combination with heat-kernel techniques, the second one being a weak
coupling expansion.

As an application of our results, in the previous section we discussed
the Bose-Einstein condensation for the case of a constant background
field. Qualitatively the properties are the same as for flat Minkowski
spacetime, however the results (\ref{7.6}) and (\ref{7.7}) show how the
critical quantities at the condensation point depend on the curvature.

The results given in this article mainly include only the leading order
correction to the thermodynamic potential at high temperature when the
chemical potentials are non-zero.
A more accurate result which includes sub-dominant terms in the
temperature can
be obtained by using the local momentum space expansion of Bunch and
Parker \cite{BunchParker} for the Green function in place of the flat
spacetime Green function (\ref{eq:8.32})
we have used. It is also possible to extend the result to the next
order in
$\mu$.
This will be given elsewhere.\\
\\
\\
\vspace{5mm}
\ni{\large \bf Acknowledgments}

It is a pleasure to thank E.~Elizalde and S.D.~Odintsov
for discussions and
remarks that turned into an improvement of the manuscript.
K.K. thanks the Department ECM of Barcelona University for
warm hospitality.
K.K. acknowledges financial support from the Alexander von Humboldt
Foundation (Germany) and by CIRIT (Generalitat de Catalunya). D.J.T. would like
to thank the Nuffield Foundation for their support.
\bs
\begin{appendix}
\renewcommand{\theequation}{{\mbox A}.\arabic{equation}}
\section{Appendix: Diagonalization of the matrix}
\setcounter{equation}{0}
The aim of this appendix is the diagonalization of the matrix $U(\cl )$.
The matrix $S$ which diagonalizes the matrix $U (\cl )$ by
$U_{diag}(\cl )=S^{-1}U(\cl )S$ is given by
\beq
S=\left(\begin{array}{cccccc}
      e_1 & e_2 & e_3 & \cdot & \cdot & e_N \\
      e_2 & -e_1 & 0 & \cdot & \cdot & 0 \\
      e_3 & 0 & -e_1  & \cdot & \cdot & 0 \\
      \cdot & \cdot & \cdot & \cdot & \cdot & \cdot \\
      \cdot & \cdot & \cdot & \cdot & \cdot & \cdot \\
      e_N & 0 & 0 & \cdot & \cdot & -e_1
      \end{array}
   \right),\label{eq:a.1}
\eeq
with the inverse
\beq
S^{-1}=\left(\begin{array}{cccccc}
      e_1 & e_2 & e_3 & \cdot & \cdot & e_N \\
      e_2 & -\frac{1-e_2^2}{e_1} & \frac{e_3}{e_1}e_2 & \cdot & \cdot &
 \frac{e_N}{e_1}e_2 \\
      e_3 & \frac{e_2}{e_1}e_3 & -\frac{1-e_3^2}{e_1}  & \cdot & \cdot &
\frac{e_N}{e_1}e_2 \\
      \cdot & \cdot & \cdot & \cdot & \cdot & \cdot \\
      \cdot & \cdot & \cdot & \cdot & \cdot & \cdot \\
      e_N & \frac{e_2}{e_1}e_N & \frac{e_3}{e_1}e_N & \cdot & \cdot &
- \frac{1-e_N^2}{e_1}
      \end{array}
   \right).\label{eq:a.2}
\eeq

With these results at hand, eqs.~(\ref{eq:2.12}) and (\ref{eq:2.18})
for $\zeta_z ^0$ and $\zeta_z ^2$ are derived. The only thing to use
furthermore is
\beq
tr\left\{\exp\left[-t\left(-U(\cl )-\frac 1 6 R\right)
\right]a_i\right\}&=&
tr\left\{S^{-1}\exp\left[-t\left(U(\cl )-\frac 1 6 R\right)
\right]SS^{-1}a_iS\right\}\nn\\
& &\hspace{-6cm}=
     tr\left\{\left(
    \begin{array}{cccc}
    \exp(-tM_1^2) & 0 & \cdot & 0 \\
       0 & \exp(-tM_2^2) & \cdot   & 0 \\
     \cdot & \cdot & \cdot & \cdot \\
     0 & 0 & \cdot & \exp(-tM_2^2)
    \end{array}
  \right)
  S^{-1}a_i S\right\}.\label{eq:a.3}
\eeq
\end{appendix}


\newpage
\begin{thebibliography}{99}
\bibitem[\dag]{kk}
Alexander von Humboldt-fellow. E-mail address:
klaus@zeta.ecm.ub.es.
\bibitem[\ddag]{djt}
E-mail address : d.j.toms@newcastle.ac.uk.
\bibitem[1]{haberprl}
H.E.~Haber and H.A.~Weldon, {\em Phys.~Rev.~Lett.} {\bf 46} (1981) 1497.
\bibitem[2]{haber}
H.E.~Haber and H.A.~Weldon, {\em Phys.~Rev.~D} {\bf 25} (1982) 502.
\bibitem[3]{kapusta}
J.I.~Kapusta, {\em Phys.~Rev.~D} {\bf 24} (1981) 426.
\bibitem[4]{bernstein}
J.~Bernstein and S.~Dodelson, {\em Phys.~Rev.~Lett.} {\bf 66} (1991)
683.
\bibitem[5]{benson}
K.M.~Benson, J.~Bernstein and S.~Dodelson, {\em Phys.~Rev.~D} {\bf 44}
(1991) 2480.
\bibitem[6]{Altaie}
M.B.~~Al'taie, {\em J.~Phys.~A} {\bf 11} (1978) 1603.
\bibitem[7]{singh}
S.~Singh and R.K.~Pathria, {\em J.~Phys.~A} {\bf 17} (1984) 2983.
\bibitem[8]{parkerzhang}
L.~Parker and Y.~Zhang, {\em Phys.~Rev.~D} {\bf 44} (1991) 2421.
\bibitem[9]{Shiraishi}
K.~Shiraishi, {\em Prog.~Theor.~Phys.} {\bf 77} (1987) 975.
\bibitem[10]{smithtoms}
J.~Smith and D.J.~Toms, unpublished.
\bibitem[11]{cognola}
G.~Cognola and L.~Vanzo, {\em Phys.~Rev.~D} {\bf 47} (1993) 4575.
\bibitem[12]{huang}
W.~Huang, {\em J.~Math.~Phys.} {\bf 35} (1994) 3594.
\bibitem[13]{doktor}
K.~Kirsten, {\em Class.~Quantum Grav.} {\bf 8} (1991) 2239.\\
K.~Kirsten, {\em J.~Phys.~A} {\bf 24} (1991) 3281.
\bibitem[14]{bose}
D.J.~Toms, {\em Phys.~Rev.~Lett.} {\bf 8} (1992) 1152.\\
D.J.~Toms, {\em Phys.~Rev.~D} {\bf 47} (1993) 2483.
\bibitem[15]{diag}
M.-H.~Lee, H.-C.~Kim and J.K.~Kim, {\em Bose-Einstein condensation for a
self-interacting theory in curved spacetime}, Preprint KAIST-CHEP-93/M4.
\bibitem[16]{fulling}
S.A.~Fulling, {\em Phys.~Rev.~D} {\bf 7} (1973) 2850.
\bibitem[17]{bernard}
C.W.~Bernard, {\em Phys.~Rev.~D} {\bf 9} (1974) 3312.
\bibitem[18]{background}
B.S.~De Witt, {\em Dynamical Theory of Groups and Fields}, (Gordon and
Breach, New York, 1965).
\bibitem[19]{haw}
S.W.~Hawking, {\em Commun.~Math.~Phys.} {\bf 55} (1977) 133.
\bibitem[20]{critch}
R.~Critchley and J.S.~Dowker, {\em Phys.~Rev.~D} {\bf 13} (1976) 3224.
\bibitem[21]{heat}
J.S.~Dowker and G.~Kennedy, {\em J.~Phys.~A} {\bf 11} (1978) 895.\\
J.S.~Dowker and J.P.~Schofield, {\em Nucl.~Phys.~B} {\bf 327} (1989)
267.
\bibitem[22]{onmodel}
E.~Elizalde, K.~Kirsten and S.D.~Odintsov, {\em Phys.~Rev.~D} {\bf 50}
(1994) 5137.
\bibitem[23]{kernel}
R.~Seeley, {\em Proc.~Symp.~Pure Math.} {\bf 10} (1967) 288.\\
R.~Seeley, {\em Am.~J.~Math.} {\bf 91} (1961) 889.\\
S.~Minakshisundaram and A.~Pleijel, {\em Can.~J.~Math.} {\bf 1} (1948)
242.
\bibitem[24]{david}
L.~Parker and D.J.~Toms, {\em Phys.~Rev.~D} {\bf 31} (1985) 353. \\
L.~Parker and D.J.~Toms, {\em Phys.~Rev.~D} {\bf 31} (1985) 2424.
\bibitem[25]{leonard}
I.~Jack and L.~Parker, {\em Phys.~Rev.~D} {\bf 31} (1985) 2439.
\bibitem[26]{klaus92}
K.~Kirsten, {\em J.~Phys.~A} {\bf 25} (1992) 6297.
\bibitem[27]{finite}
K.~Kirsten, {\em Class.~Quantum Grav.} {\bf 10} (1993) 1461.
\bibitem[28]{dolanjackiw}
L.~Dolan and R.~Jackiw, {\em Phys.~Rev.~D} {\bf 9} (1974) 3320.
\bibitem[29]{fendley}
P.~Fendley, {\em Phys.~Lett.~B} {\bf 196} (1987) 175.
\bibitem[30]{Abers}
E.S.~Abers and B.W.~Lee, {\em Phys. Rep.} {\bf 9} (1973),1.
\bibitem[31]{BunchParker}
T.S.~Bunch and L.~Parker, {\em Phys.~Rev.~D} {\bf 20} (1979) 2499.
\end{thebibliography}
\end{document}